\documentclass[cameraready]{Interspeech}
\usepackage{float}

\usepackage{booktabs}
\usepackage{caption}
\usepackage{siunitx}
\usepackage[table]{xcolor}
\definecolor{silver}{rgb}{0.75,0.75,0.75}
\usepackage{hyperref}
\usepackage{cleveref}

\usepackage{atbegshi}

\usepackage{fontawesome5}

\makeatletter
\def\ps@arxivbanner{%
  \let\@mkboth\@gobbletwo
  \def\@oddhead{\hfil
    \raisebox{0pt}[0pt][0pt]{\small\itshape Accepted at Interspeech 2026}%
    \hfil}%
  \let\@evenhead\@oddhead
  \def\@oddfoot{}\let\@evenfoot\@oddfoot
}
\makeatother

\definecolor{gold}{HTML}{FFD700}
\definecolor{silver}{HTML}{C0C0C0}
\definecolor{bronze}{HTML}{CD7F32}

\newcommand{\goldstar}{\textcolor{gold}{\faStar}}

\definecolor{gold}{HTML}{FFD700}
\definecolor{silver}{HTML}{C0C0C0}
\definecolor{bronze}{HTML}{CD7F32}

\newcommand{\silverdiamond}{\textcolor{silver}{$\blacklozenge$}}
\newcommand{\bronzetriangle}{\textcolor{bronze}{$\blacktriangle$}}

\title{CAL-MOS: Bridging Layers with Adapters for Robust MOS Prediction Across Speech Foundation Models}

\author[affiliation={1,2}, correspondingauthor]{Alef Iury}{Ferreira}
\author[affiliation={1}]{Pedro}{Lustosa Rege Botelho}
\author[affiliation={1,3}]{Fernanda}{Silva}
\author[affiliation={1,4}]{Daniel}{Casanova}
\author[affiliation={1,3}]{Rafael}{Faustino}
\author[affiliation={1,2}]{Frederico}{Oliveira}
\author[affiliation={1,2}]{Arlindo Galvão}{Filho}
\author[affiliation={1,2}]{Anderson}{da Silva Soares}

\address{
    $^1$ AKCIT, Brazil \\
    $^2$ Federal University of Goiás (UFG), Brazil \\
    $^3$ Federal University of Rio Grande do Norte (UFRN), Brazil\\
    $^4$ Federal University of Technology (UTFPR), Brazil
}

\email{alef\_iury\_c.c@discente.ufg.br}

\keywords{Speech Quality Assessment, Self-Supervised Learning, Speech Foundation Models}

\usepackage{comment}

\begin{document}

\maketitle

\thispagestyle{arxivbanner}

\begin{abstract}
Speech Quality Assessment (SQA) is essential for modern speech technologies, and recent non-intrusive SQA predictors increasingly rely on Speech Foundation Models (SFMs). However, because SFMs expose representations from many layers, it remains unclear which depths are most informative for MOS prediction and how multi-layer information should be combined reliably across backbones and datasets. We benchmark ten SFMs on four MOS datasets under three regimes: full fine-tuning, last-layer probing with a frozen encoder, and naive cross-layer weighted aggregation. We find that the best layer is strongly backbone- and dataset-dependent, and that naive weighted fusion can be unstable across settings. We further evaluate a layer-calibrated aggregation variant that applies per-layer adapters before pooling, which improves the robustness of multi-layer fusion and narrows the gap to full fine-tuning while keeping the backbone frozen.
\end{abstract}

\section{Introduction}
\label{sec:intro}

Recent progress in speech synthesis, enhancement, and related speech generation technologies has increased the need for reliable quality evaluation methods~\cite{wagner19_ssw,PERROTIN2025101747}. Although objective metrics are efficient, they often fail to reflect the nuances of human perception~\cite{wagner19_ssw,SHEN2024127471,lo19_interspeech}. As a result, subjective Mean Opinion Score (MOS) listening tests remain the standard protocol for quality assessment~\cite{Loizou2011}. However, MOS testing is notoriously expensive, time-consuming, and difficult to scale. It is also affected by annotator variability and other sources of bias, especially for high-quality modern systems~\cite{wagner19_ssw,LEMAGUER2024101577,PERROTIN2025101747}. These limitations motivate automatic, non-intrusive MOS prediction as a practical alternative.

Much of the recent progress in non-intrusive MOS prediction has been driven by large pre-trained speech encoders, or Speech Foundation Models (SFMs). Prior work has either fine-tuned the full encoder with a regression head, achieving strong performance at a higher computational cost~\cite{tseng21b_interspeech,9746395cooper2022,saeki22c_interspeech}, or trained lightweight predictors on frozen SFM representations for greater efficiency~\cite{oliveira2023mos}. Additional studies have combined SFM embeddings with complementary acoustic features to improve robustness~\cite{zezario2024mosanetplus}. Together, these results establish SFMs as strong backbones for MOS prediction, but leave open the question of how best to exploit their internal representations.

This problem arises because different layers capture fundamentally different types of information. Earlier and intermediate layers often retain stronger acoustic and phonetic cues, whereas deeper layers are increasingly shaped by the pre-training objective~\cite{pasad2021layer,pasad2023layerwise}. For MOS prediction, where cues such as noise, distortion, articulation, and prosody are all relevant, this distinction is especially important~\cite{close2023perceive}. Recent studies have indeed shown that early-to-mid layers can outperform the final layer for MOS prediction across several backbones~\cite{liang2025selection,cooper2025layer}.

Motivated by this, we revisit MOS prediction from SFMs as a practical problem of layer utilization. We benchmark ten SFMs on four MOS datasets under a unified protocol, comparing last-layer probing, best-layer selection, naive cross-layer weighted aggregation, and full fine-tuning. We further evaluate a layer-calibrated aggregation variant based on per-layer adapters, inspired by prior work in speaker verification~\cite{li2025enhancing,cai2024leveraging}, to study whether calibrating layer outputs before fusion improves the robustness of MOS prediction.

Our contributions are threefold: (1) an empirical study of layer utilization for MOS prediction across diverse speech foundation models, datasets, and training regimes, offering practical guidance on when to use the last layer, an intermediate layer, cross-layer aggregation, or full fine-tuning; (2) showing that the most informative depth is strongly backbone- and dataset-dependent, and that naive weighted-sum fusion is not a reliable general solution across backbones and datasets; and (3) showing that applying prior layer calibration before aggregation provides a stronger and more robust frozen-backbone alternative than naive weighted aggregation, often narrowing the gap to full fine-tuning.

\section{Related Works}
\label{sec:related-works}

SQA has shifted from traditional signal-based methods to data-driven deep learning systems. While early models like MOSNet~\cite{lo19_interspeech} marked significant progress, the current paradigm is dominated by SFMs models. Fine-tuning self-supervised models such as wav2vec 2.0~\cite{NEURIPS2020_92d1e1eb} was shown to surpass earlier approaches for MOS prediction~\cite{NEURIPS2020_92d1e1eb,tseng21b_interspeech}, and subsequent studies confirmed the effectiveness of other large speech encoders, including HuBERT~\cite{9746395cooper2022} and supervised models like Whisper-based~\cite{pmlr-v202-radford23a} systems~\cite{oliveira2023mos,zezario2024mosanetplus}. Stronger systems such as UTMOS~\cite{saeki22c_interspeech} further improved performance through more elaborate training pipelines, while MOSA-Net+~\cite{zezario2024mosanetplus} combined Whisper embeddings with spectral features to improve robustness.

A related line of work has examined which internal representations of pre-trained speech models are most useful for perceptual prediction. Layer-wise studies in speech representation learning show that intermediate layers often retain stronger acoustic and phonetic information, whereas deeper layers become increasingly specialized to the pre-training objective~\cite{pasad2021layer,pasad2023layerwise,close2023perceive}. In MOS prediction, recent studies by Liang et al.~\cite{liang2025selection} and Cooper et al.~\cite{cooper2025layer} similarly found that early-to-mid layers can outperform the final layer across several backbones and evaluation settings.

Our work builds most directly on this line of research, but moves beyond single-layer selection to study how multiple layers can be used robustly across backbones and datasets. This connects to prior cross-layer aggregation work in other speech tasks, particularly speaker verification, where learnable layer weighting, multi-scale aggregation, and per-layer adapters have been used to better align intermediate representations before fusion~\cite{li2025enhancing,cai2024leveraging}. Motivated by these ideas, we compare last-layer probing, best-layer selection, naive weighted aggregation, and full fine-tuning under a unified MOS benchmark and further examine whether per-layer calibration can make cross-layer fusion more reliable for MOS prediction.

\section{Methods}
\label{sec:methods}

We study non-intrusive MOS prediction using SFMs under a unified framework designed to isolate the effect of layer utilization. Given an input utterance, an SFM produces hidden representations at multiple depths. Our central question is how these layer-wise representations should be used for MOS prediction.

We evaluate a diverse set of publicly available SFMs: wav2vec 2.0 Large~\cite{NEURIPS2020_92d1e1eb}, XLS-R~\cite{babu22_interspeech} 300M and 1B, MMS~\cite{JMLR:v25:23-1318} 300M and 1B, Wav2BERT 2.0~\cite{barrault2023seamless}, WavLM Large~\cite{chen2022wavlm}, HuBERT Large~\cite{9746395cooper2022}, data2vec Large~\cite{baevski2022data2vec}, and Whisper Large-v3~\cite{pmlr-v202-radford23a}. These models differ in training objective, supervision, scale, and pre-training data, making them suitable for studying whether layer-selection trends generalize across backbones.

Our first strategy is \textit{Last Layer} (LL), where only the final hidden representation of the frozen backbone is used, serving as a standard feature-extraction baseline. Our second strategy is \textit{Best Layer} (BL), which uses a single frozen intermediate layer selected from layer-wise analysis, testing whether careful layer selection alone is sufficient. Our third strategy is \textit{Weighted Sum} (WS), where all layer representations are combined via learnable scalar weights normalized during training like done by~\cite{yang21c_interspeech}, providing a simple form of cross-layer aggregation that assumes representations from different depths are already compatible enough to be fused directly.

We also evaluate \textit{Full Fine-Tuning} (FT), where the entire SFM and regression head are optimized jointly, serving as a strong baseline. In addition, motivated by the instability observed in naive cross-layer fusion, we introduce \textit{Adapter + Mean} (A+M), a layer-calibrated aggregation strategy. Here, each layer is first passed through an independent adapter consisting of a linear projection, layer normalization, ReLU activation, and a second linear layer. The adapted sequences are then concatenated across layers and mean-pooled to form a fixed-dimensional utterance representation, allowing each layer to be transformed into a more task-aligned space before aggregation without requiring full backbone adaptation.

After layer selection or aggregation, masked mean pooling over time produces an utterance-level embedding, which is passed to a Multi-Layer Perceptron (MLP) regression head consisting of a linear projection, a ReLU activation, and a final linear output layer; predictions are clipped to the $[1, 5]$ interval to match the MOS scale. All models are trained by minimizing Mean Squared Error (MSE) between predicted and reference MOS values, with a batch size of 64 and the AdamW optimizer ($\beta_1 = 0.9$, $\beta_2 = 0.98$, $\epsilon = 1 \times 10^{-8}$), a weight decay of $1 \times 10^{-6}$, and norm-based gradient clipping (maximum threshold of 10). A cosine learning rate scheduler with a 500-step linear warm-up adjusts learning rates between $1 \times 10^{-5}$ and $5 \times 10^{-5}$.

We evaluate on four MOS datasets: BVCC~\cite{9746395cooper2022}, BRSpeechMOS~\cite{oliveira2023mos}, SingMOS~\cite{tang2024singmos}, and TMHINT-QI~\cite{chen22i_interspeech}. These datasets cover different languages, speech generation conditions, and rating scenarios. We report utterance-level and system-level results using MSE (Mean Square Error) and Spearman Rank Correlation Coefficient (SRCC). To compare models under a consistent budget, we first run a 20-epoch screening stage across all candidate backbones and training regimes to identify strong models and characterize broad trends. We then perform detailed layer-wise analysis on the most competitive backbones and extend training to 100 epochs for the final comparison.

\begin{table*}[!th]
\centering
\scriptsize
\captionsetup{skip=2pt}
\renewcommand{\arraystretch}{0.92}
\setlength{\tabcolsep}{2.0pt}

\caption{Results of the 20-epoch screening stage, comparing model families using utterance-level and system-level MSE and SRCC on BRSpeech (BRS), BVCC (BVC), SingMOS (SM), and TMHINT-QI (TMH). \textbf{Bold} indicates the best result within each model family, while \colorbox{blue!20}{blue} and \colorbox{gray!25}{gray} cell backgrounds denote the overall best and second-best entries, respectively. Avg. Utt. and Avg. Sys. report averages across the four datasets at the utterance and system levels. LL = Last Layer, WS = Weighted Sum, and FT = Fine-Tuning.}
\label{tab:is2026_20epochs_mse_srcc_compact}

\sisetup{
    reset-text-series = false,
    text-series-to-math = true,
    mode=text,
    tight-spacing=true,
    round-mode=places,
    round-precision=3,
    table-format=1.3,
    table-number-alignment=center
}

\begin{tabular}{@{}l
    *{4}{S} !{\color{gray!30}\vrule}
    *{4}{S} !{\color{gray!30}\vrule}
    *{4}{S} !{\color{gray!30}\vrule}
    *{4}{S} !{\color{gray!60}\vrule width 0.8pt}
    *{4}{S}@{}}
\toprule
& \multicolumn{4}{c}{\textbf{BRS}} & \multicolumn{4}{c}{\textbf{BVC}} & \multicolumn{4}{c}{\textbf{SM}} & \multicolumn{4}{c}{\textbf{TMH}} & \multicolumn{4}{c}{\textbf{Avg.}} \\
\cmidrule(lr){2-5}\cmidrule(lr){6-9}\cmidrule(lr){10-13}\cmidrule(lr){14-17}\cmidrule(lr){18-21}
\textbf{Method} &
\multicolumn{2}{c}{\textbf{Utt.}} & \multicolumn{2}{c}{\textbf{Sys.}} &
\multicolumn{2}{c}{\textbf{Utt.}} & \multicolumn{2}{c}{\textbf{Sys.}} &
\multicolumn{2}{c}{\textbf{Utt.}} & \multicolumn{2}{c}{\textbf{Sys.}} &
\multicolumn{2}{c}{\textbf{Utt.}} & \multicolumn{2}{c}{\textbf{Sys.}} &
\multicolumn{2}{c}{\textbf{Utt.}} & \multicolumn{2}{c}{\textbf{Sys.}} \\
\cmidrule(lr){2-3}\cmidrule(lr){4-5}
\cmidrule(lr){6-7}\cmidrule(lr){8-9}
\cmidrule(lr){10-11}\cmidrule(lr){12-13}
\cmidrule(lr){14-15}\cmidrule(lr){16-17}
\cmidrule(lr){18-19}\cmidrule(lr){20-21}
& {MSE} & {SRCC} & {MSE} & {SRCC}
& {MSE} & {SRCC} & {MSE} & {SRCC}
& {MSE} & {SRCC} & {MSE} & {SRCC}
& {MSE} & {SRCC} & {MSE} & {SRCC}
& {MSE} & {SRCC} & {MSE} & {SRCC} \\
\midrule
w2v2-L LL & 0.990 & 0.369 & 0.354 & 0.714 & 0.775 & 0.489 & 0.589 & 0.661 & 0.578 & 0.114 & 0.183 & 0.403 & 0.607 & 0.581 & 0.134 & 0.718 & 0.738 & 0.388 & 0.315 & 0.624 \\
w2v2-L WS & 0.790 & 0.592 & \textbf{0.072} & \cellcolor{gray!25}\textbf{0.943} & 0.228 & 0.856 & 0.134 & 0.898 & 0.516 & 0.381 & 0.152 & 0.562 & \textbf{0.378} & \textbf{0.743} & \textbf{0.040} & \textbf{0.935} & \textbf{0.478} & \textbf{0.643} & \textbf{0.099} & \textbf{0.835} \\
w2v2-L FT & \textbf{0.783} & \textbf{0.665} & 0.101 & \cellcolor{gray!25}\textbf{0.943} & \textbf{0.190} & \textbf{0.881} & \textbf{0.100} & \textbf{0.924} & \textbf{0.395} & \textbf{0.591} & \cellcolor{blue!20}\textbf{0.040} & \cellcolor{blue!20}\textbf{0.876} & 1.558 & 0.395 & 0.636 & 0.208 & 0.732 & 0.633 & 0.219 & 0.738 \\
\midrule
XLSR-300M LL & 1.007 & 0.454 & 0.287 & 0.657 & 0.536 & 0.661 & 0.399 & 0.718 & 0.546 & 0.208 & 0.183 & 0.152 & 0.477 & 0.686 & 0.100 & 0.834 & 0.641 & 0.502 & 0.242 & 0.590 \\
XLSR-300M WS & 0.965 & 0.579 & 0.262 & 0.886 & 0.228 & 0.858 & 0.124 & 0.914 & 0.569 & 0.407 & 0.166 & 0.655 & 0.356 & 0.752 & \textbf{0.030} & \textbf{0.940} & 0.529 & 0.649 & 0.146 & 0.849 \\
XLSR-300M FT & \textbf{0.731} & \cellcolor{gray!25}\textbf{0.693} & \textbf{0.055} & \cellcolor{gray!25}\textbf{0.943} & \cellcolor{gray!25}\textbf{0.170} & \cellcolor{blue!20}\textbf{0.894} & \textbf{0.096} & \textbf{0.935} & \textbf{0.395} & \textbf{0.585} & \cellcolor{gray!25}\textbf{0.053} & \textbf{0.819} & \textbf{0.350} & \textbf{0.762} & 0.041 & 0.924 & \cellcolor{blue!20}\textbf{0.411} & \textbf{0.733} & \cellcolor{gray!25}\textbf{0.061} & \textbf{0.905} \\
\midrule
XLSR-1B LL & 0.973 & 0.513 & 0.258 & 0.886 & 0.295 & 0.822 & 0.163 & 0.873 & 0.421 & 0.532 & 0.104 & 0.752 & 0.375 & 0.752 & 0.048 & 0.930 & 0.516 & 0.655 & 0.143 & 0.860 \\
XLSR-1B WS & 0.819 & 0.604 & 0.201 & 0.886 & 0.226 & 0.865 & 0.136 & 0.918 & 0.420 & 0.528 & 0.097 & 0.699 & 0.378 & 0.759 & 0.061 & \textbf{0.939} & 0.461 & 0.689 & 0.124 & 0.861 \\
XLSR-1B FT & \textbf{0.744} & \textbf{0.668} & \textbf{0.044} & \cellcolor{gray!25}\textbf{0.943} & \textbf{0.178} & \textbf{0.892} & \textbf{0.099} & \textbf{0.937} & \textbf{0.404} & \textbf{0.587} & \textbf{0.081} & \textbf{0.782} & \textbf{0.352} & \textbf{0.768} & \textbf{0.037} & 0.936 & \textbf{0.419} & \textbf{0.729} & \textbf{0.065} & \textbf{0.899} \\
\midrule
MMS-300M LL & 1.201 & 0.523 & 0.405 & 0.886 & 0.538 & 0.760 & 0.373 & 0.780 & 0.580 & 0.386 & 0.203 & 0.635 & 0.419 & 0.727 & 0.073 & 0.861 & 0.684 & 0.599 & 0.264 & 0.790 \\
MMS-300M WS & 0.845 & 0.554 & 0.189 & 0.886 & 0.214 & 0.864 & 0.096 & 0.921 & 0.448 & 0.491 & \textbf{0.075} & 0.721 & \cellcolor{gray!25}\textbf{0.345} & 0.765 & \textbf{0.027} & \textbf{0.946} & 0.463 & 0.669 & \textbf{0.097} & 0.869 \\
MMS-300M FT & \cellcolor{blue!20}\textbf{0.685} & \textbf{0.689} & \cellcolor{gray!25}\textbf{0.026} & \cellcolor{gray!25}\textbf{0.943} & \textbf{0.172} & \cellcolor{blue!20}\textbf{0.894} & \cellcolor{gray!25}\textbf{0.085} & \cellcolor{gray!25}\textbf{0.940} & \textbf{0.416} & \textbf{0.606} & 0.094 & \textbf{0.811} & 0.510 & \cellcolor{blue!20}\textbf{0.773} & 0.193 & 0.935 & \textbf{0.446} & \cellcolor{gray!25}\textbf{0.741} & 0.100 & \textbf{0.907} \\
\midrule
MMS-1B LL & 1.069 & 0.477 & 0.361 & 0.886 & 0.308 & 0.809 & 0.203 & 0.863 & 0.480 & 0.454 & 0.141 & 0.575 & 0.373 & 0.753 & 0.034 & 0.943 & 0.557 & 0.623 & 0.185 & 0.817 \\
MMS-1B WS & 0.788 & 0.614 & 0.120 & 0.886 & 0.237 & 0.862 & 0.137 & 0.923 & \textbf{0.403} & 0.565 & \textbf{0.070} & 0.820 & \textbf{0.349} & \textbf{0.761} & \cellcolor{gray!25}\textbf{0.027} & \textbf{0.950} & \textbf{0.444} & 0.701 & \textbf{0.089} & 0.895 \\
MMS-1B FT & \textbf{0.758} & \textbf{0.678} & \textbf{0.111} & \cellcolor{gray!25}\textbf{0.943} & \cellcolor{blue!20}\textbf{0.170} & \cellcolor{gray!25}\textbf{0.893} & \cellcolor{blue!20}\textbf{0.076} & \cellcolor{blue!20}\textbf{0.942} & 0.480 & \textbf{0.588} & 0.162 & \textbf{0.828} & 0.439 & 0.757 & 0.124 & 0.934 & 0.462 & \textbf{0.729} & 0.118 & \textbf{0.912} \\
\midrule
w2v-BERT LL & 1.430 & -0.045 & 0.618 & -0.029 & 0.486 & 0.705 & 0.332 & 0.795 & 0.718 & -0.086 & 0.254 & -0.126 & 0.469 & 0.706 & 0.082 & 0.890 & 0.776 & 0.320 & 0.322 & 0.383 \\
w2v-BERT WS & \cellcolor{gray!25}\textbf{0.694} & 0.653 & 0.025 & 0.943 & 0.231 & 0.858 & 0.138 & 0.905 & 0.410 & 0.528 & 0.085 & 0.752 & \textbf{0.348} & \textbf{0.767} & \textbf{0.028} & \cellcolor{blue!20}\textbf{0.955} & \textbf{0.421} & 0.702 & 0.069 & 0.889 \\
w2v-BERT FT & 0.703 & \textbf{0.682} & \cellcolor{blue!20}\textbf{0.011} & \cellcolor{blue!20}\textbf{1.000} & \textbf{0.199} & \textbf{0.886} & \textbf{0.115} & \textbf{0.928} & \textbf{0.386} & \textbf{0.604} & \textbf{0.054} & \textbf{0.826} & 0.446 & 0.760 & 0.059 & 0.939 & 0.433 & \textbf{0.733} & \cellcolor{blue!20}\textbf{0.060} & \cellcolor{blue!20}\textbf{0.923} \\
\midrule
WavLM-L LL & 0.793 & 0.606 & \textbf{0.133} & 0.886 & 0.273 & 0.824 & 0.165 & 0.879 & 0.516 & 0.396 & 0.165 & 0.595 & 0.375 & 0.753 & 0.041 & 0.929 & 0.489 & 0.645 & 0.126 & 0.822 \\
WavLM-L WS & 1.333 & 0.524 & 0.531 & 0.886 & 0.308 & 0.819 & 0.139 & 0.887 & 1.040 & 0.317 & 0.388 & 0.622 & 0.446 & 0.706 & 0.047 & 0.917 & 0.782 & 0.591 & 0.276 & 0.828 \\
WavLM-L FT & \textbf{0.746} & \cellcolor{blue!20}\textbf{0.703} & 0.153 & \cellcolor{gray!25}\textbf{0.943} & \textbf{0.198} & \textbf{0.878} & \textbf{0.105} & \textbf{0.931} & \cellcolor{blue!20}\textbf{0.363} & \cellcolor{gray!25}\textbf{0.622} & \textbf{0.063} & \textbf{0.823} & \textbf{0.351} & \textbf{0.767} & \textbf{0.039} & \textbf{0.936} & \cellcolor{gray!25}\textbf{0.414} & \cellcolor{blue!20}\textbf{0.742} & \textbf{0.090} & \textbf{0.908} \\
\midrule
HuBERT-L LL & 0.790 & 0.621 & 0.093 & \textbf{0.886} & 0.319 & 0.803 & 0.178 & 0.880 & 0.474 & 0.439 & \textbf{0.126} & 0.614 & 0.391 & 0.735 & 0.045 & 0.922 & 0.493 & 0.649 & 0.111 & 0.826 \\
HuBERT-L WS & 1.536 & 0.439 & 0.781 & 0.771 & 0.347 & 0.791 & 0.195 & 0.865 & 1.073 & 0.286 & 0.382 & 0.460 & 0.427 & 0.703 & \textbf{0.042} & 0.918 & 0.846 & 0.555 & 0.350 & 0.754 \\
HuBERT-L FT & \textbf{0.749} & \textbf{0.664} & \textbf{0.052} & 0.829 & \textbf{0.230} & \textbf{0.886} & \textbf{0.160} & \textbf{0.933} & \textbf{0.438} & \cellcolor{blue!20}\textbf{0.623} & 0.129 & \textbf{0.808} & \textbf{0.376} & \textbf{0.750} & 0.046 & \textbf{0.928} & \textbf{0.448} & \textbf{0.731} & \textbf{0.097} & \textbf{0.875} \\
\midrule
d2v-L LL & 1.082 & 0.463 & 0.397 & 0.829 & 0.795 & 0.292 & 0.576 & 0.451 & 0.536 & 0.271 & 0.162 & 0.292 & 0.694 & 0.532 & 0.195 & 0.571 & 0.777 & 0.390 & 0.333 & 0.536 \\
d2v-L WS & 0.968 & 0.507 & 0.283 & 0.886 & 0.498 & 0.727 & 0.368 & 0.804 & 0.493 & 0.401 & 0.152 & 0.552 & \textbf{0.450} & 0.697 & \textbf{0.068} & 0.895 & 0.602 & 0.583 & 0.218 & 0.784 \\
d2v-L FT & \textbf{0.907} & \textbf{0.614} & \textbf{0.036} & \cellcolor{gray!25}\textbf{0.943} & \textbf{0.222} & \textbf{0.863} & \textbf{0.106} & \textbf{0.922} & \textbf{0.448} & \textbf{0.557} & \textbf{0.059} & \cellcolor{gray!25}\textbf{0.837} & 0.492 & \textbf{0.753} & 0.116 & \textbf{0.950} & \textbf{0.517} & \textbf{0.697} & \textbf{0.079} & \cellcolor{gray!25}\textbf{0.913} \\
\midrule
Whisper-Lv3 LL & \textbf{0.784} & 0.625 & \textbf{0.030} & 0.886 & 0.261 & 0.836 & 0.146 & 0.889 & 0.425 & 0.544 & 0.099 & 0.767 & \cellcolor{blue!20}\textbf{0.344} & \cellcolor{gray!25}\textbf{0.768} & \cellcolor{blue!20}\textbf{0.023} & \cellcolor{gray!25}\textbf{0.954} & \textbf{0.454} & 0.693 & \textbf{0.075} & 0.874 \\
Whisper-Lv3 WS & 1.011 & 0.513 & 0.272 & 0.829 & 0.351 & 0.787 & 0.228 & 0.849 & 0.479 & 0.437 & 0.143 & 0.550 & 0.381 & 0.745 & 0.037 & 0.937 & 0.555 & 0.621 & 0.170 & 0.791 \\
Whisper-Lv3 FT & 0.809 & \textbf{0.640} & 0.041 & \cellcolor{gray!25}\textbf{0.943} & \textbf{0.201} & \textbf{0.885} & \textbf{0.133} & \textbf{0.930} & \cellcolor{gray!25}\textbf{0.378} & \textbf{0.604} & \textbf{0.061} & \textbf{0.794} & 0.494 & 0.757 & 0.156 & 0.927 & 0.470 & \textbf{0.722} & 0.098 & \textbf{0.898} \\
\bottomrule
\end{tabular}
\end{table*}

\begin{figure}[!t]
  \centering
  \includegraphics[width=\columnwidth]{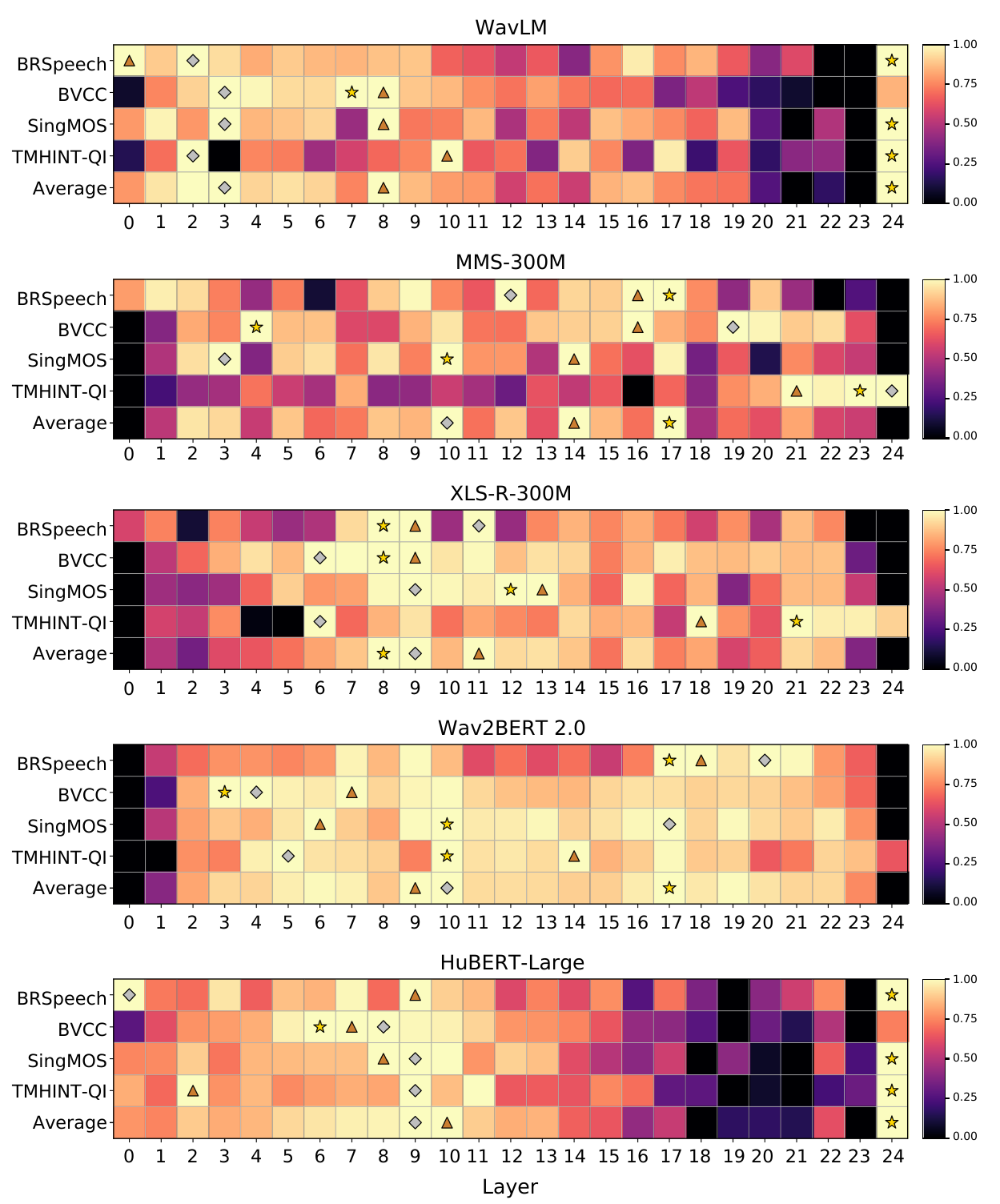}
  \caption{Layer-wise MOS prediction performance for five selected speech foundation models, reported as normalized SRCC across BRSpeech, BVCC, SingMOS, TMHINT-QI, and the average across datasets. Columns denote encoder layers and rows denote datasets. Markers indicate the top three layers in each row: best (\goldstar), second-best (\silverdiamond), and third-best (\bronzetriangle). The results show substantial variation across backbones and datasets, with strong layers often appearing in the early-to-mid portion of the network rather than consistently at the final layer.}
\label{fig:layer-wise-results}
\end{figure}

\begin{figure}[!t]
  \centering
\includegraphics[width=\columnwidth]{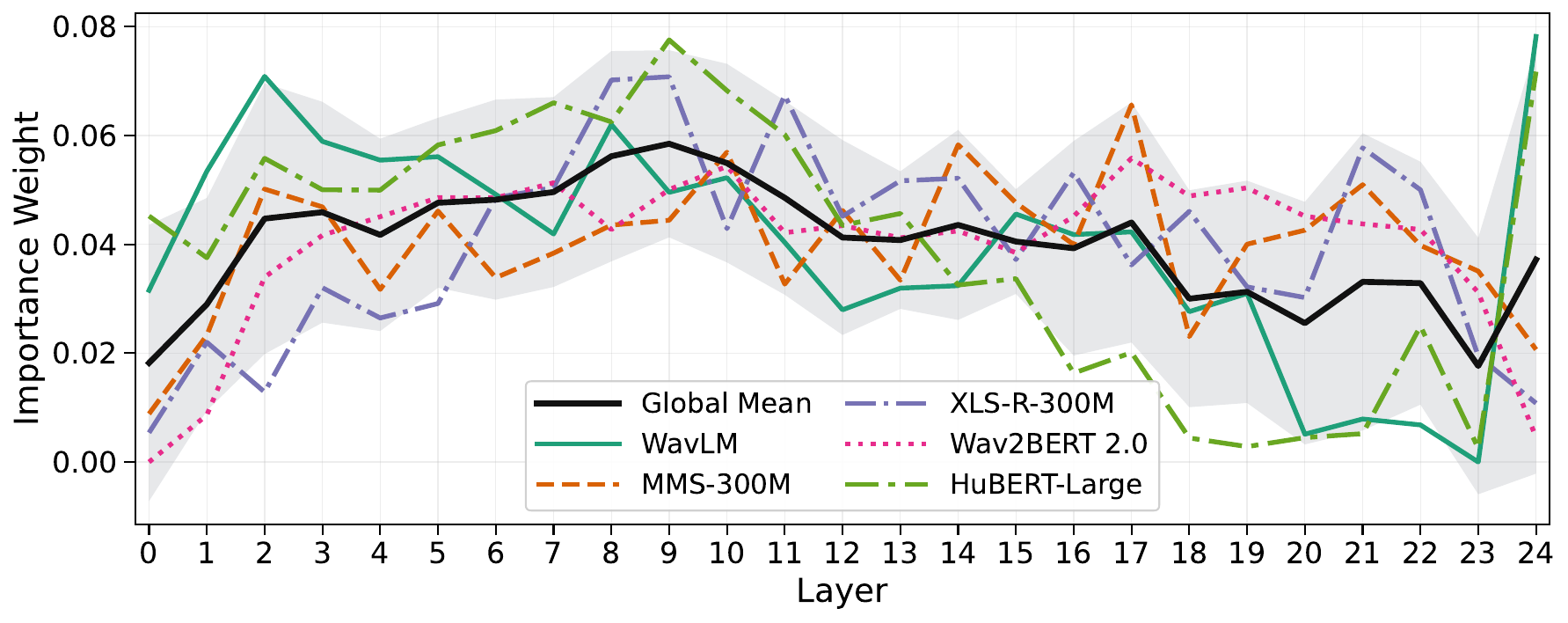}
  \caption{Mean normalized layer-importance curves for five SFMs. Colored lines denote model-specific averages, the black line shows the global mean, and the gray band indicates variability across models.}
\label{fig:layer-importance}
\end{figure}

\begin{table*}[!th]
\centering
\scriptsize
\captionsetup{skip=2pt}
\renewcommand{\arraystretch}{0.92}
\setlength{\tabcolsep}{2.0pt}

\caption{Results of the 100-epoch main comparison on the five selected model families, evaluated with utterance-level and system-level MSE and SRCC on BRSpeech (BRS), BVCC (BVC), SingMOS (SM), and TMHINT-QI (TMH). \textbf{Bold} indicates the best result within each model family, while \colorbox{blue!20}{blue} and \colorbox{gray!25}{gray} cell backgrounds denote the overall best and second-best results, respectively. Avg. Utt. and Avg. Sys. report averages across the four datasets at the utterance and system levels. LL = Last Layer, BL = Best Layer, WS = Weighted Sum, FT = Fine-Tuning, and A+M = Adapters + Mean.}
\label{tab:is2026_100epochs_filtered_mse_srcc_compact}

\sisetup{
    reset-text-series = false,
    text-series-to-math = true,
    mode=text,
    tight-spacing=true,
    round-mode=places,
    round-precision=3,
    table-format=1.3,
    table-number-alignment=center
}

\begin{tabular}{@{}l
    *{4}{S} !{\color{gray!30}\vrule}
    *{4}{S} !{\color{gray!30}\vrule}
    *{4}{S} !{\color{gray!30}\vrule}
    *{4}{S} !{\color{gray!60}\vrule width 0.8pt}
    *{4}{S}@{}}
\toprule
& \multicolumn{4}{c}{\textbf{BRS}} & \multicolumn{4}{c}{\textbf{BVC}} & \multicolumn{4}{c}{\textbf{SM}} & \multicolumn{4}{c}{\textbf{TMH}} & \multicolumn{4}{c}{\textbf{Avg.}} \\
\cmidrule(lr){2-5}\cmidrule(lr){6-9}\cmidrule(lr){10-13}\cmidrule(lr){14-17}\cmidrule(lr){18-21}
\textbf{Method} &
\multicolumn{2}{c}{\textbf{Utt.}} & \multicolumn{2}{c}{\textbf{Sys.}} &
\multicolumn{2}{c}{\textbf{Utt.}} & \multicolumn{2}{c}{\textbf{Sys.}} &
\multicolumn{2}{c}{\textbf{Utt.}} & \multicolumn{2}{c}{\textbf{Sys.}} &
\multicolumn{2}{c}{\textbf{Utt.}} & \multicolumn{2}{c}{\textbf{Sys.}} &
\multicolumn{2}{c}{\textbf{Utt.}} & \multicolumn{2}{c}{\textbf{Sys.}} \\
\cmidrule(lr){2-3}\cmidrule(lr){4-5}
\cmidrule(lr){6-7}\cmidrule(lr){8-9}
\cmidrule(lr){10-11}\cmidrule(lr){12-13}
\cmidrule(lr){14-15}\cmidrule(lr){16-17}
\cmidrule(lr){18-19}\cmidrule(lr){20-21}
& {MSE} & {SRCC} & {MSE} & {SRCC}
& {MSE} & {SRCC} & {MSE} & {SRCC}
& {MSE} & {SRCC} & {MSE} & {SRCC}
& {MSE} & {SRCC} & {MSE} & {SRCC}
& {MSE} & {SRCC} & {MSE} & {SRCC} \\
\midrule
WavLM-L LL & 0.703 & 0.686 & \textbf{0.057} & \cellcolor{gray!25}\textbf{0.943} & 0.242 & 0.849 & 0.129 & 0.910 & 0.357 & 0.608 & 0.056 & 0.835 & \textbf{0.347} & 0.764 & \textbf{0.026} & \textbf{0.946} & 0.412 & 0.727 & 0.067 & 0.908 \\
WavLM-L BL & 0.703 & 0.686 & \textbf{0.057} & \cellcolor{gray!25}\textbf{0.943} & 0.234 & 0.865 & 0.140 & 0.914 & 0.357 & 0.608 & 0.056 & 0.835 & \textbf{0.347} & 0.764 & \textbf{0.026} & \textbf{0.946} & 0.410 & 0.731 & 0.070 & 0.909 \\
WavLM-L WS & 0.868 & 0.625 & 0.077 & 0.943 & 0.270 & 0.845 & 0.140 & 0.908 & 0.446 & 0.553 & 0.048 & 0.847 & 0.407 & 0.730 & 0.042 & 0.922 & 0.498 & 0.688 & 0.077 & 0.905 \\
WavLM-L FT & 0.720 & 0.689 & 0.067 & 0.943 & \textbf{0.188} & \textbf{0.885} & \textbf{0.098} & \textbf{0.930} & 0.376 & 0.602 & 0.050 & 0.849 & 0.384 & \textbf{0.772} & 0.082 & 0.943 & 0.417 & 0.737 & 0.074 & 0.916 \\
WavLM-L A+M & \cellcolor{gray!25}\textbf{0.691} & \cellcolor{blue!20}\textbf{0.698} & 0.072 & 0.943 & 0.189 & 0.883 & 0.103 & 0.925 & \cellcolor{gray!25}\textbf{0.342} & \cellcolor{gray!25}\textbf{0.636} & \cellcolor{gray!25}\textbf{0.041} & \textbf{0.864} & \textbf{0.347} & 0.772 & 0.046 & 0.934 & \cellcolor{gray!25}\textbf{0.392} & \cellcolor{gray!25}\textbf{0.747} & \textbf{0.065} & \textbf{0.917} \\
\midrule
MMS-300M LL & 0.759 & 0.621 & 0.078 & 0.886 & 0.248 & 0.839 & 0.142 & 0.884 & 0.423 & 0.510 & 0.094 & 0.649 & 0.345 & 0.764 & 0.026 & 0.948 & 0.444 & 0.684 & 0.085 & 0.842 \\
MMS-300M BL & 0.996 & \textbf{0.684} & 0.374 & \textbf{0.943} & 0.211 & 0.878 & 0.137 & 0.914 & \textbf{0.380} & 0.611 & \textbf{0.050} & \textbf{0.841} & 0.345 & 0.764 & 0.029 & 0.945 & 0.483 & 0.734 & 0.148 & \textbf{0.911} \\
MMS-300M WS & \textbf{0.750} & 0.655 & \textbf{0.044} & 0.829 & 0.207 & 0.876 & 0.128 & 0.924 & 0.407 & 0.574 & 0.074 & 0.780 & 0.359 & 0.757 & 0.042 & 0.925 & \textbf{0.431} & 0.716 & \textbf{0.072} & 0.865 \\
MMS-300M FT & 0.821 & 0.652 & 0.102 & \textbf{0.943} & \cellcolor{blue!20}\textbf{0.169} & \cellcolor{blue!20}\textbf{0.894} & \cellcolor{blue!20}\textbf{0.072} & \cellcolor{blue!20}\textbf{0.944} & 0.428 & 0.554 & 0.096 & 0.749 & \textbf{0.339} & \cellcolor{blue!20}\textbf{0.775} & \cellcolor{gray!25}\textbf{0.025} & \cellcolor{gray!25}\textbf{0.954} & 0.439 & 0.719 & 0.074 & 0.897 \\
MMS-300M A+M & 0.805 & 0.679 & 0.124 & \textbf{0.943} & 0.251 & 0.888 & 0.202 & 0.930 & 0.486 & \textbf{0.611} & 0.156 & 0.787 & 0.401 & 0.771 & 0.084 & 0.952 & 0.486 & \textbf{0.737} & 0.142 & 0.903 \\
\midrule
XLSR-300M LL & 0.793 & 0.612 & 0.098 & 0.886 & 0.281 & 0.818 & 0.174 & 0.872 & 0.433 & 0.508 & 0.113 & 0.669 & 0.367 & 0.747 & 0.032 & 0.949 & 0.469 & 0.671 & 0.104 & 0.844 \\
XLSR-300M BL & 0.734 & \textbf{0.687} & \cellcolor{gray!25}\textbf{0.039} & 0.943 & 0.198 & 0.875 & 0.105 & 0.916 & 0.371 & 0.630 & 0.060 & \cellcolor{gray!25}\textbf{0.864} & 0.364 & 0.754 & 0.030 & 0.948 & 0.417 & 0.737 & 0.058 & 0.918 \\
XLSR-300M WS & \textbf{0.722} & 0.674 & 0.075 & 0.943 & 0.202 & 0.876 & 0.104 & 0.923 & 0.385 & 0.607 & 0.058 & 0.815 & 0.360 & 0.760 & 0.043 & 0.948 & 0.417 & 0.729 & 0.070 & 0.907 \\
XLSR-300M FT & 0.735 & 0.678 & 0.055 & \cellcolor{blue!20}\textbf{1.000} & 0.173 & 0.892 & \cellcolor{gray!25}\textbf{0.082} & \textbf{0.940} & 0.384 & 0.594 & 0.054 & 0.799 & 0.337 & 0.770 & \cellcolor{blue!20}\textbf{0.023} & \textbf{0.953} & \textbf{0.407} & 0.733 & \cellcolor{gray!25}\textbf{0.053} & \textbf{0.923} \\
XLSR-300M A+M & 0.771 & 0.686 & 0.148 & 0.943 & \cellcolor{gray!25}\textbf{0.172} & \cellcolor{gray!25}\textbf{0.893} & 0.090 & 0.933 & \textbf{0.352} & \textbf{0.631} & \textbf{0.046} & 0.826 & \cellcolor{gray!25}\textbf{0.337} & \textbf{0.772} & 0.028 & 0.950 & 0.408 & \textbf{0.745} & 0.078 & 0.913 \\
\midrule
Wav2BERT LL & 0.752 & 0.638 & 0.162 & 0.771 & 0.302 & 0.801 & 0.155 & 0.883 & 0.445 & 0.449 & 0.117 & 0.659 & 0.366 & 0.752 & 0.029 & 0.954 & 0.466 & 0.660 & 0.116 & 0.817 \\
Wav2BERT BL & \cellcolor{blue!20}\textbf{0.679} & 0.685 & \textbf{0.042} & 0.943 & 0.200 & 0.880 & 0.127 & 0.927 & 0.355 & 0.612 & 0.046 & 0.867 & 0.346 & 0.762 & 0.029 & 0.944 & 0.395 & 0.735 & 0.061 & 0.920 \\
Wav2BERT WS & 0.699 & 0.686 & 0.062 & 0.943 & 0.198 & 0.878 & 0.106 & 0.929 & 0.347 & 0.625 & 0.044 & 0.847 & 0.341 & 0.769 & \textbf{0.026} & \cellcolor{blue!20}\textbf{0.956} & 0.396 & 0.739 & 0.059 & 0.919 \\
Wav2BERT FT & 0.827 & 0.674 & 0.151 & \cellcolor{blue!20}\textbf{1.000} & 0.188 & 0.886 & 0.109 & 0.928 & 0.440 & 0.545 & 0.073 & 0.791 & 0.375 & 0.757 & 0.047 & 0.905 & 0.458 & 0.716 & 0.095 & 0.906 \\
Wav2BERT A+M & 0.699 & \cellcolor{gray!25}\textbf{0.691} & 0.050 & 0.943 & \textbf{0.178} & \textbf{0.890} & \textbf{0.092} & \textbf{0.934} & \cellcolor{blue!20}\textbf{0.341} & \cellcolor{blue!20}\textbf{0.644} & \cellcolor{blue!20}\textbf{0.039} & \cellcolor{blue!20}\textbf{0.902} & \cellcolor{blue!20}\textbf{0.335} & \cellcolor{gray!25}\textbf{0.773} & 0.026 & 0.949 & \cellcolor{blue!20}\textbf{0.388} & \cellcolor{blue!20}\textbf{0.749} & \cellcolor{blue!20}\textbf{0.052} & \cellcolor{blue!20}\textbf{0.932} \\
\midrule
HuBERT-L LL & \textbf{0.691} & 0.685 & 0.032 & 0.943 & 0.254 & 0.841 & 0.123 & 0.912 & \textbf{0.377} & 0.587 & 0.073 & 0.787 & 0.367 & 0.748 & 0.035 & \textbf{0.940} & 0.422 & 0.715 & 0.066 & 0.895 \\
HuBERT-L BL & \textbf{0.691} & 0.685 & 0.032 & 0.943 & 0.216 & 0.872 & 0.149 & 0.896 & \textbf{0.377} & 0.587 & 0.073 & 0.787 & 0.367 & 0.748 & 0.035 & \textbf{0.940} & \textbf{0.413} & 0.723 & 0.072 & 0.891 \\
HuBERT-L WS & 0.806 & 0.657 & \cellcolor{blue!20}\textbf{0.028} & \cellcolor{blue!20}\textbf{1.000} & 0.259 & 0.855 & 0.151 & 0.918 & 0.467 & 0.519 & \textbf{0.051} & 0.844 & 0.407 & 0.726 & \textbf{0.032} & 0.938 & 0.485 & 0.689 & 0.065 & \cellcolor{gray!25}\textbf{0.925} \\
HuBERT-L FT & 0.746 & 0.677 & 0.052 & 0.943 & \textbf{0.191} & \textbf{0.888} & 0.101 & \cellcolor{gray!25}\textbf{0.942} & 0.382 & 0.605 & 0.061 & 0.811 & 0.385 & 0.754 & 0.065 & 0.921 & 0.426 & 0.731 & 0.070 & 0.904 \\
HuBERT-L A+M & 0.692 & \textbf{0.690} & 0.030 & 0.943 & 0.209 & 0.878 & \textbf{0.100} & 0.920 & 0.394 & \textbf{0.623} & 0.071 & \textbf{0.856} & \textbf{0.360} & \textbf{0.760} & 0.047 & 0.931 & 0.414 & \textbf{0.738} & \textbf{0.062} & 0.913 \\
\bottomrule
\end{tabular}
\end{table*}

\section{Results and Discussion}
\label{sec:results}

Table~\ref{tab:is2026_20epochs_mse_srcc_compact} first provides a 20-epoch screening across ten SFMs and three training regimes. A clear pattern is that performance depends strongly on both the backbone and the layer-utilization strategy. FT is a strong baseline overall, but frozen-backbone approaches remain competitive when layers are used appropriately. At the same time, the screening results already show that naive WS is not uniformly reliable: it is competitive for some backbones, such as XLSR-300M, MMS-300M, and Wav2BERT, but clearly degrades for others, notably WavLM-L, HuBERT-L, and Whisper-Lv3. This is an important early finding: simply using and aggregating more layers does not guarantee better MOS prediction.

The screening stage was also used to select the backbones for the subsequent layer-wise study and the 100-epoch comparison in Table~\ref{tab:is2026_100epochs_filtered_mse_srcc_compact}. We retained WavLM-L, HuBERT-L, XLSR-300M, MMS-300M, and Wav2BERT because they were the strongest model families overall in terms of average SRCC in the 20-epoch screening. We also preferred the 300M multilingual variants over the corresponding 1B models because the larger models did not show a sufficiently clear SRCC advantage to justify the substantially higher cost of exhaustive per-layer analysis and longer training.

The layer-wise heatmaps in Fig.~\ref{fig:layer-wise-results} and the mean importance curves in Fig.~\ref{fig:layer-importance} help explain these trends. Across models, the strongest layers are often found in the early-to-mid portion of the encoder rather than consistently at the final layer. The global mean importance curve shows the same tendency. However, the figures also make clear that there is no universal best depth: the strongest layer varies across backbones and datasets, and some models exhibit broad high-performing regions while others show sharper peaks and valleys. This explains both why LL can be suboptimal and why WS can be unstable: the most useful information is distributed differently across depth and is not necessarily aligned well enough for direct fusion.

The 100-epoch results confirm that better layer selection often matters more than simply defaulting to the final layer. The clearest gains from BL over LL appear for XLSR-300M and Wav2BERT. For XLSR-300M, the average utterance-level MSE decreases from 0.469 to 0.417 and the average system-level MSE from 0.104 to 0.058. For Wav2BERT, the corresponding improvements are from 0.466 to 0.395 and from 0.116 to 0.061. These results reinforce the main conclusion of the layer-wise analysis: the final layer is often not the strongest default representation for MOS prediction. For WavLM-L and HuBERT-L, however, BL brings smaller gains, indicating that the usefulness of intermediate layers is backbone-dependent rather than universal.

A second key finding is that naive cross-layer aggregation is not a reliable substitute for careful layer selection. WS remains competitive for Wav2BERT and reasonably strong for MMS-300M and XLSR-300M, but it is clearly weaker for WavLM-L and HuBERT-L. In these cases, directly combining hidden states from different depths with scalar weights appears too restrictive. The issue is not only which layers are useful, but also whether their representations are compatible enough to be fused without adaptation.

This is where A+M strategy becomes most useful. Across the final comparison, A+M is the strongest and most consistent frozen-backbone aggregation method, and the best overall configuration is Wav2BERT A+M, which reaches 0.388/0.749 in average utterance-level MSE/SRCC and 0.052/0.932 in average system-level MSE/SRCC. The figures suggest a plausible reason for this behavior: Wav2BERT exhibits a relatively broad band of strong middle-to-late layers rather than a single narrow optimum. This makes it a good candidate for calibrated multi-layer fusion, since several layers appear informative but not identical. In this setting, the adapters can help reshape these useful representations before pooling, making aggregation more effective than either LL, BL, or naive WS.

WavLM-L shows a different but still favorable pattern. Its heatmap suggests that useful information is present at multiple depths, but with a more irregular structure, including weaker late-layer regions. This likely makes naive WS vulnerable to mixing strong and weak representations. Consistent with this interpretation, WavLM-L A+M is the second-best overall configuration and improves markedly over WS, while also remaining slightly stronger than FT on the average utterance-level metrics. HuBERT-L behaves similarly, although the gains from A+M are more moderate. By contrast, XLSR-300M already has a relatively stable high-performing region, so BL is already strong and the gains from A+M are smaller. MMS-300M is the main mixed case: its strongest layers shift more across datasets, and A+M improves some ranking metrics but is less consistently favorable on MSE, suggesting stronger cross-dataset variation in where useful information resides.

Overall, the results support three main conclusions. First, MOS-relevant information is distributed across the depth of speech foundation models, and the best-performing layer is strongly backbone- and dataset-dependent. Second, naive weighted fusion is not a reliable general solution for exploiting multiple layers. Third, lightweight per-layer calibration makes cross-layer aggregation substantially more robust, with the largest benefits appearing when several layers are informative but not trivially compatible, as seen most clearly for Wav2BERT and WavLM-L.

\section{Conclusion}
\label{sec:conclusion}

This work revisited non-intrusive MOS prediction from speech foundation models through the lens of layer utilization. Across ten backbones and four datasets, we showed that the final layer is often not the most informative representation for MOS prediction, with strong performance frequently emerging in early-to-mid layers instead. We also found that naive cross-layer weighted fusion is not a reliable general solution, as its effectiveness depends strongly on the backbone. To address this, we introduced a layer-calibrated aggregation strategy based on per-layer adapters followed by pooling, which consistently improves robustness of multi-layer fusion and yields strongest overall frozen-backbone results, most notably with Wav2BERT and WavLM-L. Overall, our findings suggest that progress in MOS prediction depends not only on stronger backbones, but also on using their hierarchical representations more effectively. Future work will explore alternative compact fusion designs and evaluate generalization under broader cross-domain settings.

\section{Use of Generative AI Disclosure}
The authors used generative AI tools, specifically large language models, only to polish the writing of portions of the manuscript. These tools were used for minor editing tasks such as improving grammar, clarity, and phrasing. The authors take full responsibility for the content of the manuscript.

\section{Acknowledgements}

This work has been fully/partially funded by the project 
Research and Development of Algorithms for Construction of Digital Human Technological Components supported by Advanced Knowledge Center in Immersive Technologies (AKCIT), with financial resources from the PPI IoT of the MCTI grant number 057/2023, signed with EMBRAPII. The authors also wish to acknowledge the Artificial Intelligence Lab at Recod.ai and the Institute of Computing, University of Campinas, for granting access to the computational infrastructure required for the experiments.

\bibliographystyle{IEEEtran}
\bibliography{refs}

\end{document}